\documentclass[a4paper]{jpconf}
\usepackage{graphicx}
\begin{document}
\title{Basic properties of three-leg Heisenberg tube}

\author{Satoshi Nishimoto}

\address{Leibniz-Institut f\"ur Festk\"orper- und Werkstoffforschung 
Dresden, D-01171 Dresden, Germany}

\ead{s.nishimoto@ifw-dresden.de}

\author{Mitsuhiro Arikawa}

\address{Institute of Physics, University of Tsukuba 1-1-1 Tennodai, 
Tsukuba Ibaraki 305-8571, Japan}

\ead{arikawa@sakura.cc.tsukuba.ac.jp}

\begin{abstract}
We study three-leg antiferromagnetic Heisenberg model with the periodic 
boundary conditions in the rung direction. Since the rungs form regular 
triangles, spin frustration is induced. We use the density-matrix renormalization 
group method to investigate the ground state. We find that the spin excitations 
are always gapped to remove the spin frustration as long as the rung coupling 
is nonzero. We also visibly confirm spin-Peierls dimerization order in the leg 
direction. Both the spin gap and the dimerization order are basically enhanced 
as the rung coupling increases.
\end{abstract}

\section{Introduction}

For many years spin ladder systems have attracted much attention. 
The fundamental properties are essentially understood when the open boundary 
conditions are applied in the rung direction. One of the overriding properties 
is that spin-$\frac{1}{2}$ ladders are gapful for an even number of legs and 
whereas gapless for an odd number of legs (e.g., as a review, see Ref.~[1]). 
However, it has been recognized that the spin states of odd-leg ladders 
are drastically changed by applying the periodic boundary conditions in 
the rung direction (referred as a spin tube). At present, there are some 
experimental candidates for odd-leg spin tubes~\cite{Millet99,Seeber04}. 
Theoretically, it was suggested that all the spin excitations of three-leg 
Heisenberg tube are gapped due to a frustration-induced spin-Peierls 
transition~\cite{Schulz96,Kawano97}. Although further several theoretical 
studies~\cite{Sakai05,Cabra97,Cabra98,Citro00,Sato07,Luscher04,Okunishi05,Fouet06} 
have been carried out since then, the basic properties of odd-leg spin tube is 
a still open issue.

\section{Model}

\begin{figure}[t]
\begin{center}
    \includegraphics[width= 8.0cm,clip]{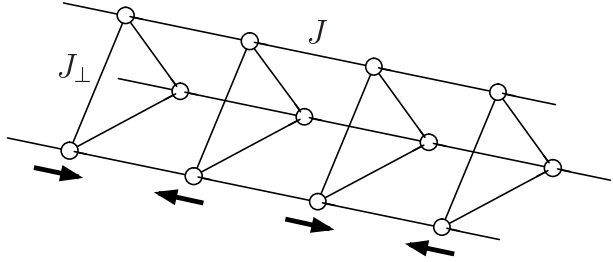}
  \caption{
Lattice structure of three-leg Heisenberg tube. The arrows show an example of 
spin-Peierls distortion.
  }
\end{center}
    \label{lattice}
\end{figure}

In general, the low-energy physics of any odd-leg spin tube may be epitomized by 
that of the three-leg spin tube. Therefore, we consider the three-leg 
antiferromagnetic Heisenberg tube, the Hamiltonian of which 
is given by
\begin{eqnarray}
H = J \sum_{\alpha=1}^3 \sum_i \vec{S}_{\alpha,i} \cdot \vec{S}_{\alpha,i+1} 
+ J_\perp \sum_{\alpha (\neq \alpha^\prime)} \sum_i \vec{S}_{\alpha,i} 
\cdot \vec{S}_{\alpha^\prime,i},
\label{hamiltonian}
\end{eqnarray}
where $\vec{S}_{\alpha,i}$ is a spin-$\frac{1}{2}$ operator at rung $i$ and leg $\alpha$. 
$J$ ($>0$) is the exchange interaction in the leg direction and $J_\perp$ ($>0$) is 
the exchange interaction between the legs (see Figure 1). We take $J=1$ as the unit of energy hereafter. 

\section{Physical quantities}

In this work we employ the density-matrix renormalization group (DMRG) method 
which provides very accurate data for ground-state properties of one-dimensional 
quantum systems; for a review, see Refs.~[14]. We use the DMRG method to 
calculate the spin gap $\Delta_\sigma$ and the dimerization order parameter $D$. 
We study ladders with several kinds of length $L = 24$ to $312$ with open-end 
boundary conditions in the leg direction. We keep up to $m=2400$ density-matrix 
eigenstates in the renormalization procedure and extrapolate the calculated 
quantities to the limit $m \to \infty$. In this way, the maximum truncation error, 
i.e., the discarded weight, is less than $1 \times10^{-7}$, while the maximum error 
in the ground-state energy is less than $10^{-7}-10^{-6}$. 

The spin gap is evaluated by an energy difference between the first triplet excited 
state and the singlet ground state,
\begin{equation}
\Delta_\sigma(L)=E(L,1)-E(L,0), \ \ \ \Delta_\sigma=\lim_{L \to \infty}\Delta_\sigma(L),
\end{equation}
where $E(L,S_z)$ is the ground-state energy of a system of length $L$, 
i.e., $L \times 3$ ladder, with $z$-component of the total spin $S_z$. 
Note that the number of system length must be taken as $L=2l$, with $l (>1)$ being 
an integer to maintain the total spin of the ground state as $S=0$. All values 
of the spin gap shown in this paper are extrapolated to the thermodynamic limit 
$L \to \infty$.

Let us then define the dimerization order parameter. Since the translational 
symmetry is broken due to the Friedel oscillation under the application of the 
open-end boundary conditions, the dimerized state is directly observable. 
We are interested in the formation of alternating spin-singlet pairs in the leg 
direction, so that we calculate the nearest-neighbor spin-spin correlations,
\begin{equation}
S(i) = -\left\langle \vec{S}_{\alpha,i} \cdot \vec{S}_{\alpha,i+1} \right\rangle,
\end{equation}
where $\left< \cdots \right>$ denotes the ground-state expectation value. 
The results for all $\alpha$ values are equivalent. It is generally known that 
the Friedel oscillations at the center of the system decay as a function of 
the system length. If the amplitude at the center of the system persists 
for arbitrarily long system length, there exists a long-ranged order. 
It corresponds to the spin-Peierls dimerized ground state in our model. 
We thus define the dimerization order parameter as 
\begin{equation}
D = \left|S(L/2) - S(L/2+1)\right|.
\end{equation}
We confirmed that $D$ is almost saturated at $L \ge 120$ in our previous paper~\cite{na3leg}, 
so that in this paper it is calculated for a system with length $L=120$. 
Nonzero value of $D$ indicates a long-ranged spin-Peierls state with finite spin gap.

\section{Results}

\begin{figure}[t]
\begin{center}
    \includegraphics[width= 6.5cm,clip]{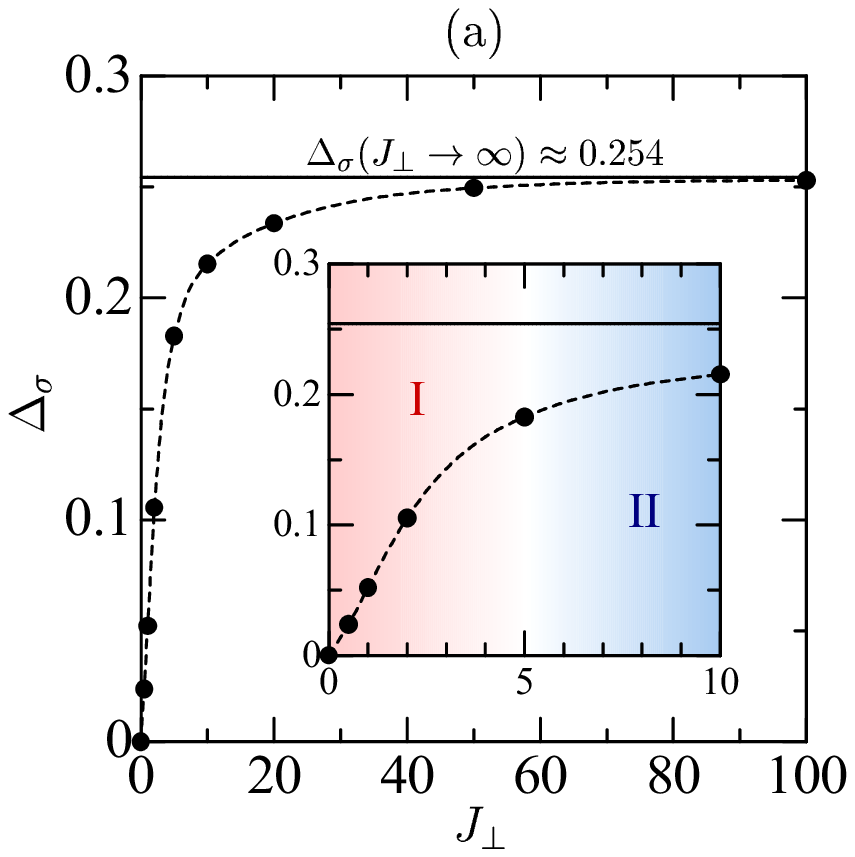}
    \includegraphics[width= 6.6cm,clip]{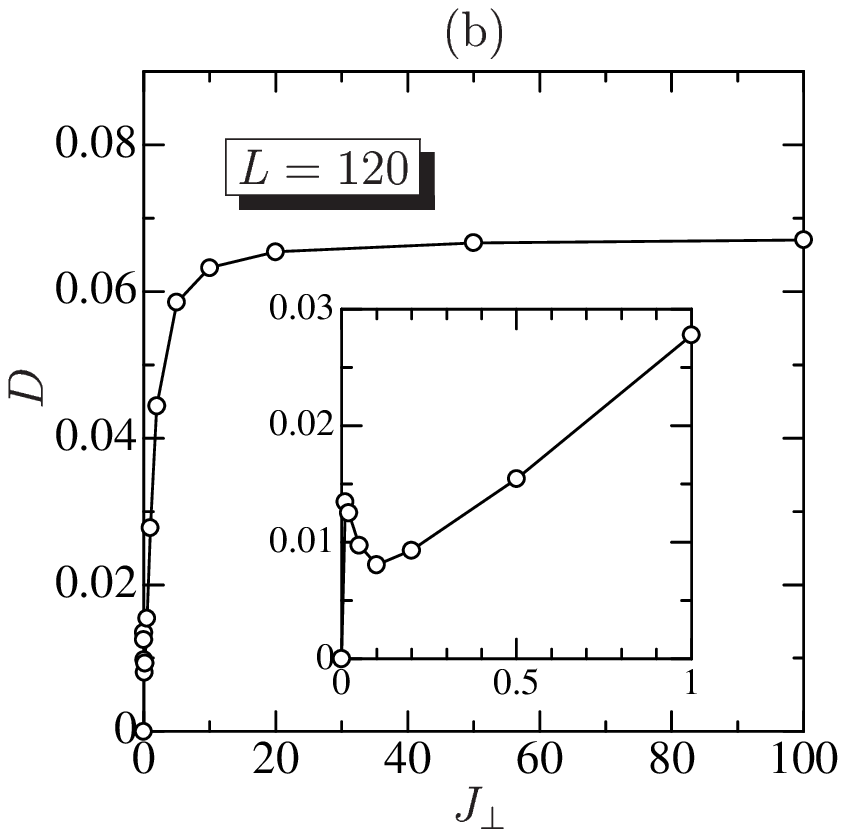}
  \caption{
(a) Extrapolated values of spin gap to the thermodynamic limit $L \to \infty$ 
as a function of $J_\perp$. The inset shows an expanded view for 
$0 \le J_\perp \le 10$. Indices I and II denote on-leg and on-rung regions, 
respectively (see text). 
(b) Dimerization order parameter for $L=120$ as a function of $J_\perp$. 
The inset shows an expanded view for $0 \le J_\perp \le 1$.
  }
\end{center}
    \label{results}
\end{figure}

\subsection{Spin gap}

In Figure 2 (a), we show the DMRG results of the spin gap $\Delta_\sigma$ as 
a function of $J_\perp$. We obtain $\Delta_\sigma=0.254$ in the limit of $J_\perp=\infty$. 
It is rather smaller than a value estimated in Ref.~[3] because of the different finite-size-scaling 
analysis. % ari
Roughly speaking, the spin gap increases proportionally to $J_\perp$ 
in the small $J_\perp$ ($\le3$) regime and keeps almost constant in the large $J_\perp$ 
($\ge10$) regime. This behavior can be interpreted in terms of different origin of 
the lowest singlet-triplet excitation for each the $J_\perp$ regime, although 
the mechanism of gap opening is invariant for the entire $J_\perp$ regime. 
Thing is, the spin gap is approximately scaled by a binding energy of most 
weakly bounded spin-singlet pair in the system and it switches around $J_\perp \approx 5$. 
In other words, most weakly bounded pairs are transferred from on-leg ones 
in the small $J_\perp$ region (on-leg region) to on-rung ones in the large $J_\perp$ 
region (on-rung region) [in the inset of Figure 2 (a), we denote the two regions 
as I and II, respectively]. A more concrete description is given in the following paragraph.

For $J \ll J_\perp$, we can easily imagine that the on-rung spin-singlet pairs must 
be bounded more solidly than the on-leg ones. The spin gap is therefore scaled by 
the binding energy of an on-leg pair, i.e., $\Delta_\sigma \propto J$. Accordingly, 
$\Delta_\sigma$ is independent of $J_\perp$ and it is consistent with the constant 
behavior of $\Delta_\sigma$ with $J_\perp$ at $J_\perp \ge 10$. 
On the other hand, the situation is somewhat different for $J_\perp < {\cal O}(J)$: 
the bound state of the on-leg pairs is expected to be more solid than that of 
the on-rung ones. It is because that the system is strongly dimerized even with 
infinitesimally small $J_\perp$. The dimerization strength develops abruptly at 
$J_\perp=0^+$ and increases rather slowly with increasing $J_\perp$ (see below). 
Thus, the spin gap is essentially scaled by the binding energy of an on-rung pair. 
In addition, we may assume that the binding energy of the on-rung pair is proportional 
to $J_\perp$ in the small $J_\perp$ regime, by analogy with that of the two-leg 
Heisenberg system~\cite{Gogolin98}. 
Now therefore, the spin gap is scaled by $J_\perp$, i.e., $\Delta_\sigma \propto J_\perp$, 
which is consistent to a linear behavior of $\Delta_\sigma$ with $J_\perp$ at 
$J_\perp \le 3$. Note that the derivative $\partial \Delta_\sigma/\partial J_\perp$ 
is very small ($\sim 0.053$) due to strong spin frustration among the intra-ring spins. 
Consequently, a crossover between the constant $\Delta_\sigma$ region and 
the proportional $\Delta_\sigma$ region is seated not at $J_\perp \approx 1$ but 
around $J_\perp \approx 5$. The existence of this crossover has also be confirmed by 
studying the $J_\perp$ dependence of the dynamical spin structure factor~\cite{na3leg}.

\subsection{Dimerization order parameter}

We plot the DMRG results of the dimerization order parameter $D$ as a function of 
$J_\perp$ in Figure 2 (b), where the system size is fixed at $L = 120$. 
We expect the $J_\perp$-dependence of $D$ to be similar to that of the spin gap 
$\Delta_\sigma$ because the binding energy of spin-singlet pairs would be scaled with 
the dimerization strength. It is true that the overall behavior seems to be similar 
to that of the spin gap. However, surprisingly, the dimerization order parameter is 
discontinuously enhanced when $J_\perp$ is switched on as contrasted with 
the linear increase of the spin gap. Then, the dimerization order parameter goes 
through a minimum around $J_\perp=0.1$ and increases almost linearly from 
$J_\perp \approx 0.2$ to $5$. In the limit of $J_\perp \to \infty$, the dimerization 
order parameter is saturated to $D \sim 0.0676$.

\section{Summary}

We study three-leg antiferromagnetic Heisenberg tube with the DMRG method. 
The spin gap and the dimerization order parameter are estimated as a function 
of the rung coupling. We suggest that the spin gap is scaled by the binding energy 
of the on-rung spin-singlet pair in the weak-coupling regime ($J_\perp \le 3$); 
whereas, it is scaled by the binding energy of the on-leg spin-singlet pair in the 
strong-coupling regime ($J_\perp \ge 10$). Furthermore, we find that 
the dimerization order parameter is approximately proportional to the spin gap 
except when the rung coupling is very small. The dimerization strength is abruptly 
enhanced at $J_\perp=0$.

\section*{Acknowledgments}

This work has been supported in part by the University of Tsukuba  
Research Initiative and Grants-in-Aid for Scientific Research,
No. 20654034 from JSPS and No. 220029004 (physics of new quantum 
phases in super clean materials) and 20046002 (Novel States of Matter 
Induced by Frustration) on Priority Areas from MEXT for MA.


\section*{References}
\begin{thebibliography}{9}
\bibitem{Dagotto96} Dagotto E and Rice T M 1996 Science {\bf 271} 618; Dagotto~E 1999
Repts. Prog. Phys. {\bf 62} 1525.

\bibitem{Millet99} Millet P, Henry J Y, Mila F, and Galy J 1999 
J. Solid State Chem. {\bf 147} 676.
\bibitem{Seeber04} Seeber G, Kogerler P, Kariuki B M, and Cronin L 2004
Chem. Commun. (Cambridge) {\bf 2004} 1580.

\bibitem{Schulz96} Schulz H J 1996 {\it Correlated Fermions and Transport in Mesoscopic Systems}, 
edited by T. Martin, G. Montambaux and T. Tr\^an Thanh V\^an (Editions Frontiers, 
Gif-sur-Yvette, France) 1996  p. 81.
\bibitem{Kawano97} Kawano K and Takahashi M 1997 J. Phys. Soc. Jpn. {\bf 66} 4001.

\bibitem{Sakai05} Sakai T, Matsumoto M, Okunishi K, Okamoto K, and Sato M 2005 
Physica E {\bf 29} 633; Sakai T, Sato M, Okunishi K, Otsuka Y, Okamoto K, Itoi C, 
arXiv:0807.4769v1.

\bibitem{Cabra97} Cabra D C, Honecker A, and Pujol P 1997 Phys. Rev. Lett. {\bf 79} 5126.
\bibitem{Cabra98} Cabra D C, Honecker A, and Pujol P 1998 Phys. Rev. B {\bf 58} 6241.
\bibitem{Citro00} Citro R, Orignac E, Andrei N, Itoi C, and Qin S, 
J. Phys.:Condens. Matter {\bf 12}, 3041 (2000).
\bibitem{Sato07} Sato M and Sakai T 2007 Phys. Rev. B {\bf 75} 014411; 
Sato M 2007 Phys. Rev. B {\bf 75} 174407.

\bibitem{Luscher04} L\"uscher A, Noack R M, Misguich G, Kotov V N, and Mila F 2004
Phys. Rev. B {\bf 70}, 060405(R).
\bibitem{Okunishi05} Okunishi K, Yoshikawa S, Sakai T, Miyashita S 2005 Prog. Theor. Phys. 
Suppl. {\bf 159} 297.
\bibitem{Fouet06} Fouet J B, L\"auchli A, Pilgram S, Noack R M, and Mila F 2006
Phys. Rev. B {\bf 73} 014409.

\bibitem{White92} White S R 1992 Phys. Rev. Lett. {\bf 69} 2863; 1993
Phys. Rev. B {\bf 48} 10345.

\bibitem{na3leg} Nishimoto S and Arikawa M 2008 Phys. Rev. B {\bf 78} in press; 
preprint arXiv:0806.2474v1.

\bibitem{Gogolin98} Gogolin A O, Nersesyan A A and Tsvelik A M, {\it Bosonization and Strongly Correlated Systems} (Cambridge University Press, Cambridge) 1998.
\end{thebibliography}
\end{document}